\documentclass[12pt]{spieman}  % 12pt font required by SPIE;
\usepackage{amsmath,amsfonts,amssymb}
\usepackage{graphicx}
\usepackage{setspace}
\usepackage{tocloft}

\title{Mie scattering due to tissue structures in the terahertz regime: Experimental and Monte Carlo verification using diffused polarimetric imaging in highly attenuating tissue phantoms}

\author[a]{Erica Heller}
\author[a]{Kuangyi Xu}
\author[a]{Zachery Harris}
\author[a,*]{M. Hassan Arbab}
\affil[a]{Stony Brook University, Department of Biomedical Engineering, 100 Nicolls Road, Stony Brook, NY, United States, 11794}

\cftpagenumbersoff{figure}
\cftpagenumbersoff{table} 
\begin{document} 
\maketitle

\begin{abstract}

\noindent
\textbf{Significance:}
Changes in the structure of tissue occur in many disease processes, such as the boundaries of cancerous tumors and burn injuries. Spectroscopic and polarimetric alterations of terahertz light caused by Mie scattering patterns has the potential to be a diagnostic marker.

\noindent
\textbf{Aim:}
We present an analysis of Monte Carlo simulation of Mie scattering of polarized terahertz light from cancerous tumor budding, compare the simulation to experimental results obtained in phantom models, and present an analysis of a polarization-sensitive terahertz scan of an ex vivo porcine burn injury. 

\noindent
\textbf{Approach:}
 Using a Monte Carlo simulation, we modeled the changes in diffuse intensity and degree of polarization of broadband off-specular terahertz light due to scattering particles in highly attenuating tissue. We extracted the Mueller matrix of the tissue using this model and analyzed its Lu-Chipman product decomposition matrices. We compared this model to experimental data from four phantoms consisting of polypropylene particles of varying sizes embedded in gelatin. Finally, we induced a full-thickness burn injury in ex vivo porcine skin samples and compared experimental data from burned and healthy regions of the tissue. 
 
 \noindent
\textbf{Results:} 
Simulation revealed contrast in the Stokes vectors and Mueller Matrix elements for varying scattering particle sizes. Experimental phantom results showed contrast between different sizes of scattering particles in degree of polarization and diffuse intensity in agreement with Monte Carlo simulation results. Finally, we demonstrated a similar diffused imaging signal contrast between burned and healthy regions of ex vivo porcine skin.

\noindent
\textbf{Conclusion:} 
Polarimetric terahertz imaging has the potential to detect structural changes due to biological disease processes. 
 
\end{abstract}

% Include a list of up to six keywords after the abstract

% Include email contact information for corresponding author
{\noindent \footnotesize\textbf{*}M. Hassan Arbab,  \linkable{hassan.arbab@stonybrook.edu} }

\begin{spacing}{2}   % use double spacing for rest of manuscript

\section{Introduction}
\label{sect:intro}  % \label{} allows reference to this section

With rapid advancement of terahertz (THz) spectroscopic methods, this range of electromagnetic frequencies has attracted more attention in the biomedical optics field in recent years \cite{leitenstorfer2023, chen:22review}. Several biological applications of THz imaging have been explored, including ex vivo experiments on cancer diagnostics \cite{ashworth_terahertz_2009,el-shenawee_cancer_2019,fan_multimodal_2017,sim_terahertz_2013}, and in vivo studies on skin burn classification \cite{KhaniJBO22,osman_deep_2022, Khani:2023:BOE:Triage}. Typically, these studies have explored the dielectric permittivity of the tissue in order to examine sample properties and to differentiate between healthy and diseased specimen. Particularly, most THz in vivo applications focus on the water content of different tissue types. In comparison, very few studies have used the scattering of the THz light from biological tissue as a diagnostic modality. With an increase in SNR and advancements in THz technology, scattering measurements could play an important role in characterization and classification of some tissue types. 
\par
Study of the dielectric permittivity of tissue often assumes a homogeneous structure, or alternatively employs an effective medium theory. This approach can be oversimplifying in many disease conditions. In particular, at the edges of certain types of cancerous tumors such as colon cancer, oral squamous cell carcinoma, and breast cancer, the formation of tumor budding and poorly differentiated clusters can occur, where parts of the tumor break off and form small clusters within otherwise healthy tissue \cite{lugli_recommendations_2017}. It has been shown that these features can be independent prognostic factors for lymph node metastasis and patient survival \cite{venkatesh_cell_2019,reggiani_bonetti_poorly_2016}. Current methods for determining the extent of growth of these features include hematoxylin and eosin (H\&E) staining and immunohistochemistry. %However, some tumor buds and poorly differentiated clusters may be obscured by inflammation, or otherwise can be difficult to distinguish from other cell types using H\&E staining, and immunohistochemistry can also stain apoptotic bodies and cellular debris 
\cite{Lugli_tumour_budding}. Other disease conditions can also create a change in the scattering properties of tissue. For example, burns can cause a destruction of higher-scattering features such as hair follicles and sweat glands in the skin. 
\par
A promising method for studying these tissue media is through the detection of polarimetric signal contrast in scattered light. Xu and Arbab recently developed a model for Mie scattering of broadband terahertz light in biological media using Monte Carlo methods \cite{Xu:24}. Light-scattering models have been used extensively in biophotonics to evaluate the potential for various modalities of optical imaging to detect different forms of skin \cite{LVWangCR2012}, gastric \cite{nishizawa_depth_2022}, breast \cite{AltoeCCR21, Mahadevan-Jansen2010}, prostate \cite{Bigio2013} and cervical cancer \cite{Richards-KortumJBO2006} subtypes, and to investigate contrast mechanisms and sensitivity to tissue changes with disease \cite{SteelmanOptica2019}. Also, polarized scattering light spectroscopy has been investigated for use in endoscopic and laparoscopic imaging of various cancers, including colonic polyps \cite{BigioColon2015} and peritoneal lesions \cite{trout_polarization_2022}. Similarly, polarization measurements of the reflected THz light have been explored for delineation between cancerous and healthy tissue \cite{doradla_detection_2013,gurjar_polarimetry_2024}; however, the mechanisms behind the polarization response differences between the types of tissue are yet to be studied in THz frequencies. Given the recent development of various THz ellipsometry and polarimetry systems \cite{Chen:18, Xu:23, Harris24, castro2005polarization, morris2012polarization,bulgarevich2014polarization}, there is a timely need for computational models that can explain the observed phenomena. 

\par
In this paper, we present Monte Carlo simulation and experimental phantom as well as ex vivo measurements of polarization changes due to Mie scattering from structures within tissue, indicating different degrees of disease progression. We explore the potential utility of diffuse scattering measurements for the THz polarimetric diagnosis of  disease processes and other THz biophotonics applications. 

\section{Methods}
\subsection{Terahertz diffused scattering spectral measurement}
\begin{figure}[ht]
\begin{center}
\begin{tabular}{c}
\includegraphics[height=8.5cm]{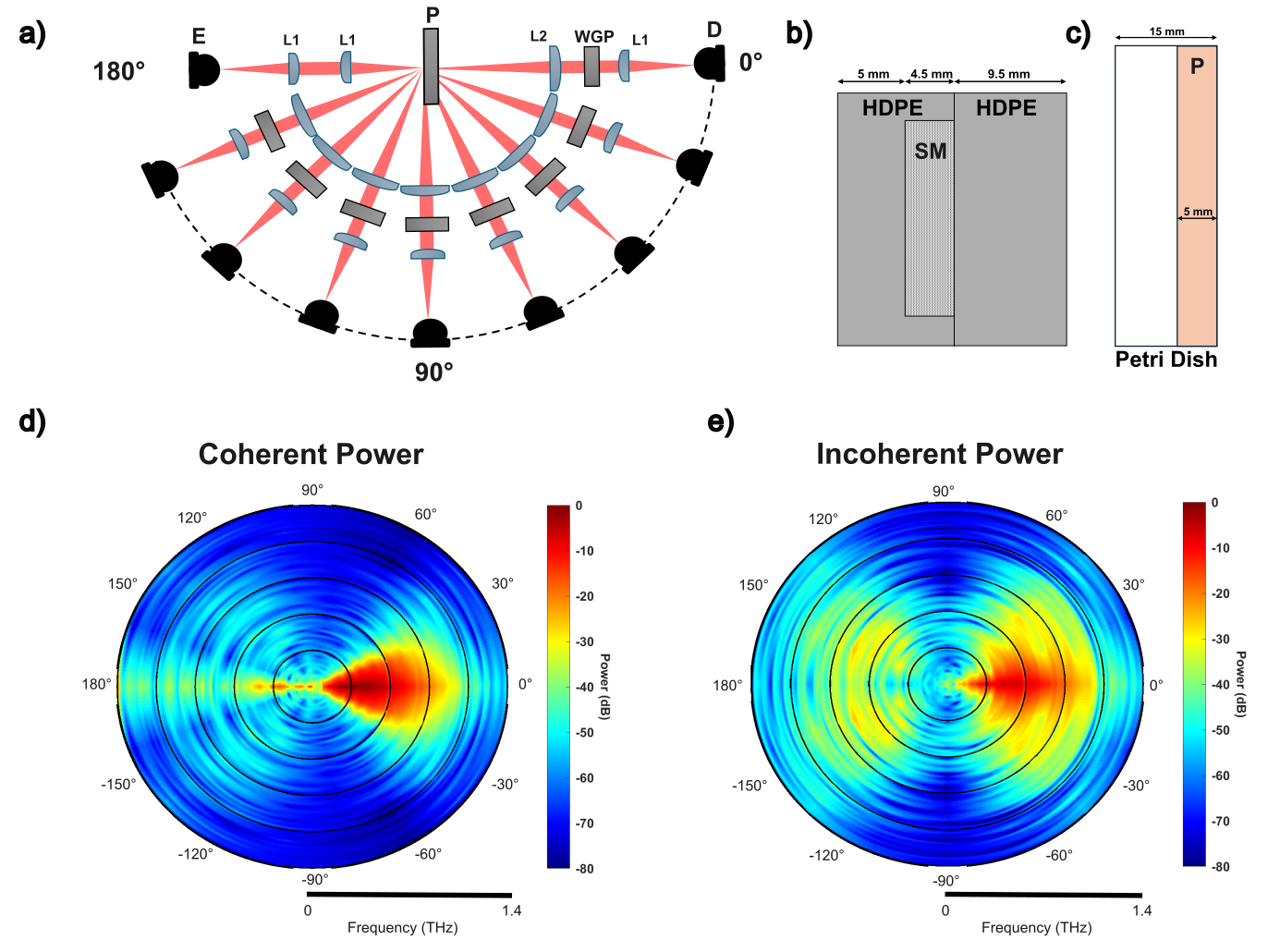}
\end{tabular}
\end{center}
\caption 
{ \label{fig:experimentalsetup}
(a) Diagram of the experimental setup. E = Emitter, D = Detector, L1 = TPX lens with 50 mm focal length, L2 = PTFE lens with 100 mm focal length, WGP = Wire-Grid Polarizer, and P = Phantom. (b) HDPE container used for holding scattering material (SM). (c) Phantom design within petri dish. (d-e) Angular distribution of coherent and incoherent power measured from LDPE scattering material. Measurement from 160-180 degrees were made with the addition of a beam-splitter in the beam path.} 
\end{figure}

Broadband terahertz time-domain spectroscopic (THz-TDS) measurements were performed using the experimental setup shown in Fig. \ref{fig:experimentalsetup}(a). In the emission arm, a photoconductive antenna (PCA) was excited using a 1560 nm femtosecond laser, which was then focused onto the sample using a pair of collimating and focusing lenses (L1). In the detection arm, which had the ability to rotate from 0-160 degrees, the light passes through a wire-grid polarizer (WGP) between the collimating (L2) and focusing lenses (L1) before it reached the PCA detector. To quantify the diffused component of the THz scattered light, we used a sample composed of 180-micron polyethylene scattering particles within a polyethylene container with a volume density of 0.3 g/cm$^3$ (Fig. \ref{fig:experimentalsetup}(b)). We obtained THz-TDS scattering measurements at 10-degree intervals from 0-160 degrees. Separately, we captured a 180-degree scattering measurement by placing a beam splitter in the emission path, and scaling the measurement according to the loss of power through the beam splitter. For each angle, scattering measurements were collected at 36 locations on the sample container, each 5 mm apart, with 100 time-domain trace averages to separate the coherent and incoherent components of the signal. Figure \ref{fig:experimentalsetup}(c) shows the phantom sample that will be used in the later experiments. Figure \ref{fig:experimentalsetup}(d-e) shows the angular distribution of the detected coherent and incoherent power, respectively, from 0.2 to 1.4 THz, normalized by the peak value of the coherent power. We observed a sharp decrease in coherent power from 0-20 degrees, with an increase past 160 degrees. In contrast, the incoherent power remained above our system’s detectable threshold (greater than -60dB) at all angles except 70-120 degrees. These results show that diffusely back-scattered beam at about 140 degrees is dominated by the incoherent scattered photons that can carry the information regarding tissue structures. Due to this sharp contrast between the incoherent and coherent power in the diffuse backscattering direction, we chose to obtain the subsequent phantom measurements at 140 degrees. These scattering data were measured by raster scanning the phantom sample over the beam focus, moving a distance of 2mm for each pixel, for x and y-polarization measurements. Two raster scans were performed for each sample, with the polarization state of the wire-grid polarizer and THz detector rotated between each scan. A total of 100 measurements were recorded for each polarization state over a 2 cm by 2 cm area. 

\subsection{Fabrication of tissue phantoms with scattering particles}
Our objective is to emulate a tissue sample, in which spherical structures having a lower refractive index and absorption coefficient than the medium are embedded. Examples of such structures include hair follicles, empty sweat glands, or certain cancer types such as oral squamous cell carcinoma \cite{THzOralSCC, THzOralSCC_2}. We constructed such phantoms using polypropylene particles embedded in gelatin. A 15\% bovine skin gelatin solution (Sigma Aldrich, Inc.) was heated to 40° C. Once the desired temperature was reached, polypropylene particles (MicroPowders, Inc.) were rapidly mixed into the solution. The solution was then poured into a Petri dish to a height of about 5mm. Due to the rapid viscosity change as the gelatin cools, the particles remained suspended within the gelatin and did not settle. This process was repeated for 4 phantoms, with differing particle sizes and concentrations. However, due to clumping of the particles during the mixing process, causing the phantoms to differ from the intended parameters, the size and concentrations of the particles were later analyzed using optical microscopy and ImageJ software \cite{ImageJ}. 

\begin{figure}[ht]
\begin{center}
\begin{tabular}{c}
\includegraphics[height=8cm]{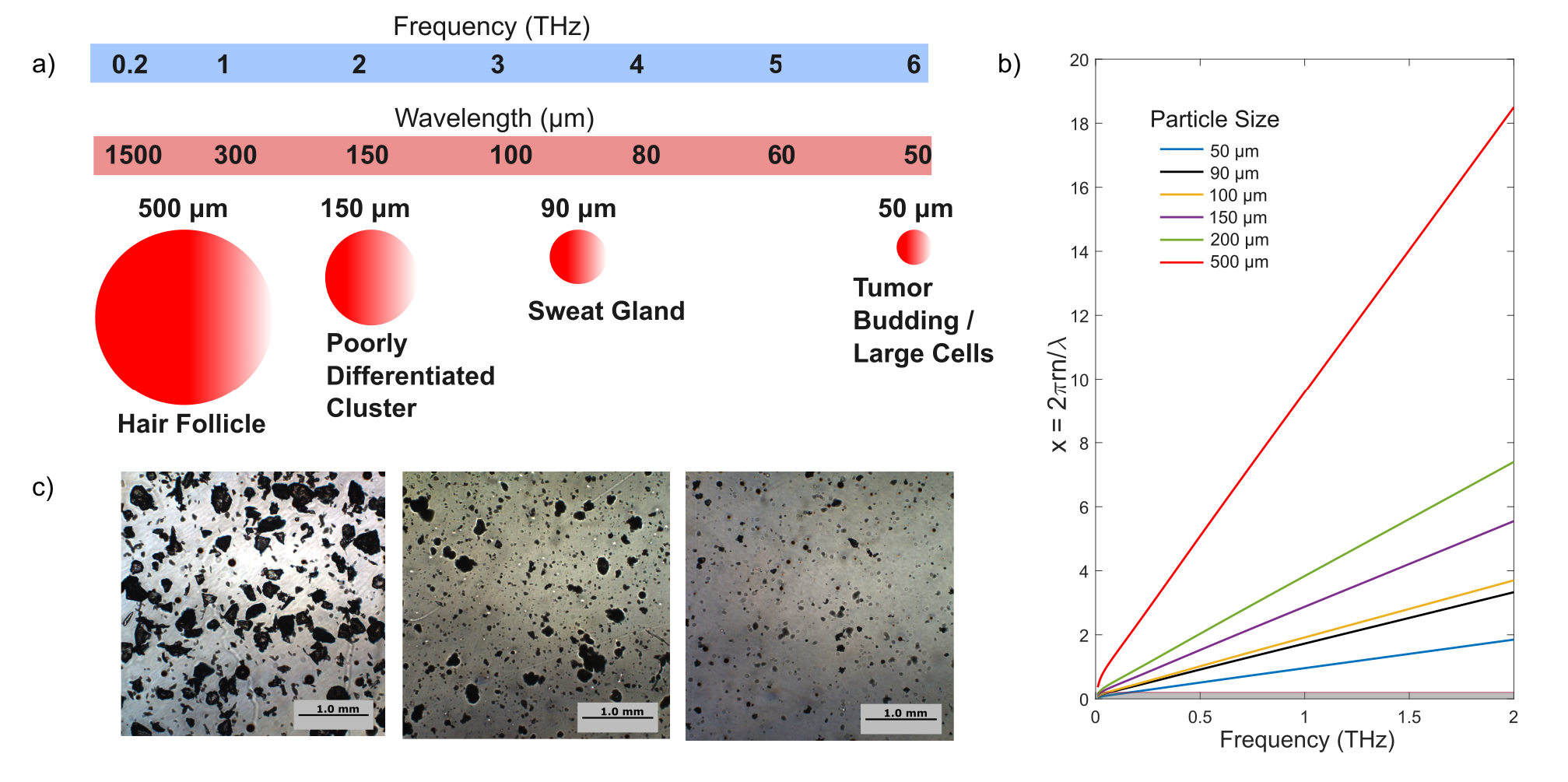}
\end{tabular}
\end{center}
\caption 
{ \label{fig:miescattering}
a) Artistic representation of relevant spherical tissue structure sizes compared to Terahertz wavelengths \cite{buddingcellnestsize}. b) Mie scattering parameter for particle sizes 50-500 $\mu$m. c) Representative optical microscopy images of fabricated phantoms. 
} 
\end{figure}

Particle sizes for the phantoms were chosen based on the wavelengths of THz light and the size of relevant tissue structures, such as tumor buds and hair follicles. Figure \ref{fig:miescattering}(a) shows a diagram of the sizes of relevant tissue structures compared to the THz wavelengths. To ensure the resultant electromagnetic scattering is within the Mie regime, we calculated the size parameter x=2$\pi$rn\slash $\lambda$ where r is the radius of the particles, n is the refractive index of the host medium (assumed to be the refractive index of skin), and lambda is the wavelength for relevant particle sizes. As shown in Fig. \ref{fig:miescattering}(b), all relevant particle sizes fall within a size parameter between 0.2-20 within our system's usable bandwidth of 0.2-1.5 THz, indicating that the primary scattering will be due to Mie theory. Particle sizes for the phantoms were chosen based on the sizes of sweat glands and poorly differentiated tumor clusters  to be reasonably detected with our experimental setup. Figure \ref{fig:miescattering}(c) shows examples of three of our phantoms, with varying particle sizes. 

\par
\begin{figure}[ht]
\begin{center}
\begin{tabular}{c}
\includegraphics[height=7.5cm]{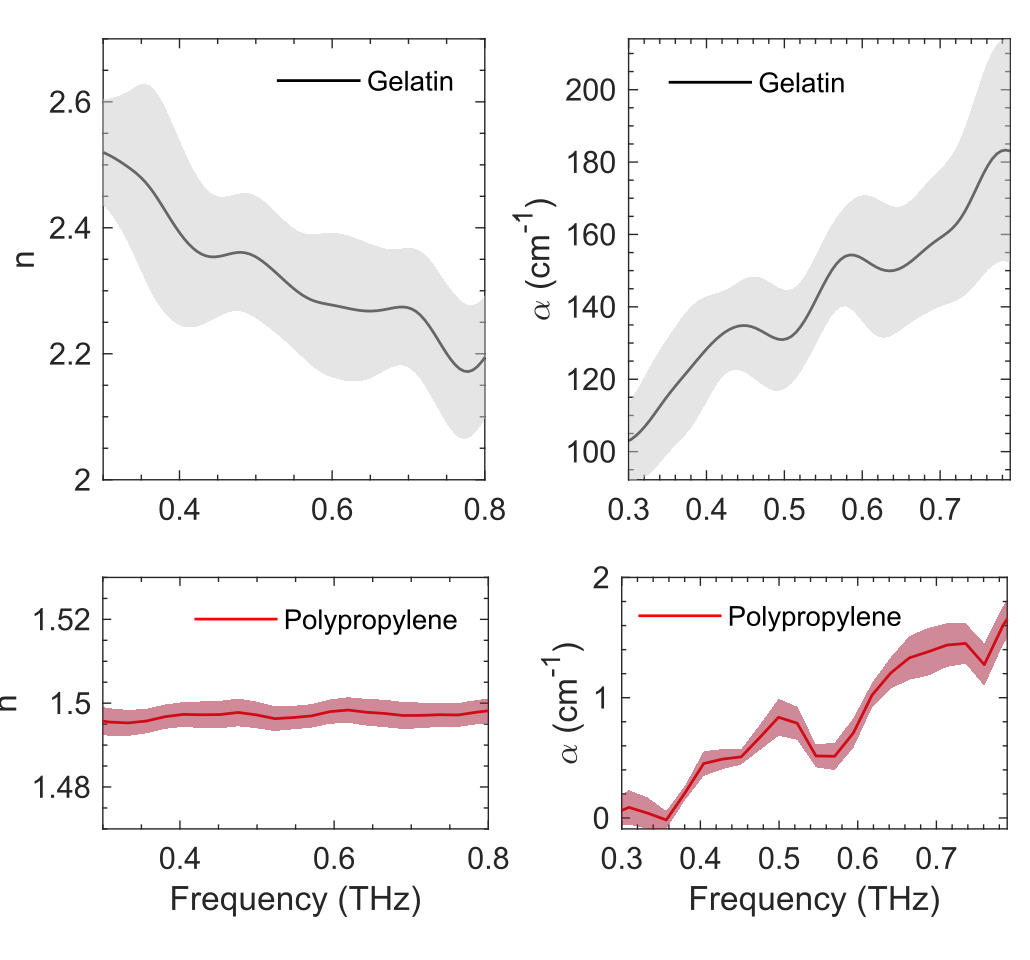}
\end{tabular}
\end{center}
\caption 
{ \label{fig:refractiveindex}
Measured refractive indexes and absorption coefficients of gelatin and polypropylene. 
} 
\end{figure}

We measured the dielectric properties of the gelatin and polypropylene particles for use in our Monte Carlo model. In the case of the polypropylene particles, we took transmission measurements at the 0-degree angle of our setup through a pellet constructed by pressing the polypropylene particles under a 3000 psi load for approximately 1 hour. For the gelatin medium, the recently developed PHASR scanner \cite{9864064, Harris24} was used to obtain reflection measurements at normal incidence angle on a pure 15\% gelatin sample. The calculated index of refraction and absorption coefficients were averaged across 100 pixels for both the particles and gelatin, shown in Fig. \ref{fig:refractiveindex}. The measured refractive index of the polypropylene particles is slightly lower than the expected value of $\sim$1.51 in the THz frequency range \cite{chang_terahertz_2020, jin_terahertz_2006}; however, this could be due to manufacturing variability or remaining air pockets within the pressed pellet sample. The measured absorption coefficient had a value of \textless 2 cm$^{-1}$ for the entire bandwidth of our system. The measured refractive index and absorption coefficient of gelatin were higher than those of skin, but below those of water.
\par

\subsection{Monte Carlo simulations}
\begin{figure}[ht]
\begin{center}
\begin{tabular}{c}
\includegraphics[height=5.5cm]{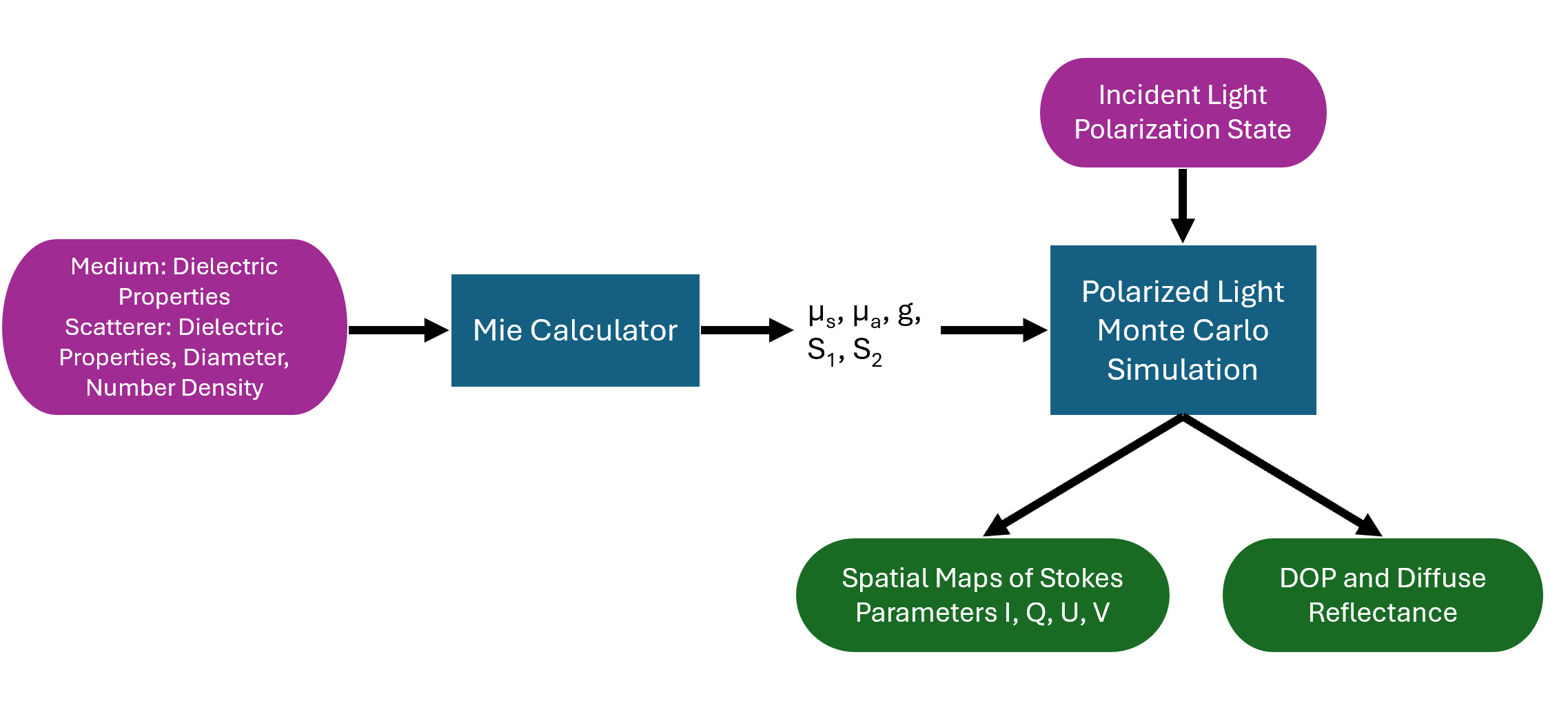}
\end{tabular}
\end{center}
\caption 
{ \label{fig:flowchart}
Flowchart indicating the simulation input parameters and the interface between the Mie Calculator and Polarized Light Monte Carlo model.   
} 
\end{figure}

The multi-later diffused reflectance and polarization Monte Carlo code used for modeling the samples was developed by Xu and Arbab \cite{Xu:24}, based on the Meridian Plane Polarized Light Monte Carlo Code by Ramella-Roman et al. \cite{ramella-roman_three_2005}, with alterations to the Mie scattering parameters to account for an absorbing host medium. Figure \ref{fig:flowchart} describes the flowchart of the altered Polarized Light Monte Carlo code with the addition of a Mie Calculator to generate the input parameters. The Mie Calculator takes an input of the dielectric properties of the medium and particles, the size of the particles, and the number density of the particles and produces the asymmetric factor g (the average cosine of the scattering angle) and the scattering and absorption coefficients $\mu_s$ and $\mu_a$, given by \cite{yang_inherent_2002}
\begin{equation}
    \mu_s = \frac{2\pi\rho}{|n_m|^2k_0^2}\sum_{j=1}^{\infty}(2j+1)(|a_j|^2+|b_j|^2)
\end{equation}

\begin{equation}
    \mu_a = \frac{2\pi\rho}{\Re(n_m)k_0^2}\Im[(\sum_{j=1}^{\infty}(2j+1)(|c_j|^2\psi_j(z)\psi_j'^*(z)-|d_j|^2\psi_j'(z)\psi_j^*(z)))/n_p]
\end{equation}

where $n_m$ and $n_p$ are the complex refractive indexes of the medium and particles respectively, $k_0 = 2\pi$/$\lambda$ is the wave vector, $\rho$ is the number density of the particles $a_j$, $b_j$, $c_j$, and $d_j$ are the Mie coefficients, and $\psi_j(z)$ is a Ricatti-Bessel function where $z=n_pk_0r$. In addition to these parameters, the Mie Calculator also produces the scattering matrix that represents the relationship between the incident ($E_i$) and scattered ($E_s$) field amplitudes using Jones calculus:

\begin{equation}
    \begin{bmatrix}
        E_s^\perp \\
        E_s^{||}
    \end{bmatrix}
    \propto
    \begin{bmatrix}
        S_1 & 0\\
        0 & S_2
    \end{bmatrix}
    \begin{bmatrix}
        E_i^\perp \\
        E_i^{||}
    \end{bmatrix}.
\end{equation}

The Mueller calculus transform of Equation 3 is given by

\begin{equation}
    \begin{bmatrix}
        I_s \\
        Q_s \\
        U_s \\
        V_s \\
    \end{bmatrix}
    \propto
    \begin{bmatrix}
        S_{11} & S_{12} & 0 & 0 \\
        S_{12} & S_{11} & 0 & 0 \\
        0 & 0 & S_{33} & S_{34} \\
        0 & 0 & -S_{34} & S_{33}
    \end{bmatrix}
    \begin{bmatrix}
        I_i \\
        Q_i \\
        U_i \\
        V_i \\
    \end{bmatrix},
\end{equation}
where I, Q, U, and V are the Stokes parameters of the incident and scattered light. The Polarized Light Monte Carlo code produces the spatial Stokes Parameters for a given set of outputs from the Mie Calculator and an initial incident light polarization. Using four different incident light polarizations (specifically horizontal linearly polarized, vertical linearly polarized, 45\textdegree \space linearly polarized, and right circularly polarized) the spatial Mueller Matrix parameters can be calculated. Additionally, using spatially averaged Stokes parameters from a linearly polarized incident light, the degree of polarization (DOP) of the scattered beam can be calculated as,

\begin{equation}
    DOP = \frac{\sqrt{Q^2+U^2+V^2}}{I}.
\end{equation}

\section{Results}

\subsection{Evaluation of average particle size and density in phantoms}

Particle sizes and concentrations for each phantom were evaluated using optical microscopy and an ImageJ workflow. A threshold was set in ImageJ to discard any detected particles below a diameter of 75$\mu$m, to eliminate any image artifacts and small particles that were unlikely to contribute to the signal. Twenty-five microscope images such as those in Fig. \ref{fig:miescattering}(c) were taken of each phantom, and 10 ROIs of 1mm x 1mm field of view were delineated for each image, giving a total of 250 measurements of average particle size and concentration for each phantom. These measurements were averaged to give an average particle size and concentration for each phantom as inputs for the simulation. The concentration calculated using this procedure provides the surface density in units of particles/mm$^2$. To convert this quantity to volume density, it was divided by the summation of the depth of field of the microscope and twice the particle diameter size. Due to inhomogeneities within the phantoms, such as clumping of the particles, some phantoms produces inconsistent density and particle size numbers and were therefore excluded from further analysis. Table \ref{table:2} summarizes the average particle sizes and volume density for the final four phantoms as input parameters for Monte Carlo analysis in the following section.

\par% Since the simulation took an input of number density, the concentration was converted from a value in particles/mm$^2$ derived from the microscope images to a number density in mm$^{-3}$ by dividing the concentration by a factor calculated by the addition of the depth of field of the microscope and 2 times the diameter of the particles to account for any particles that were partially in the field of view. 

%Due to inhomogeneities within the phantoms, three phantoms were excluded from further analysis. 

%Microscope analysis of phantoms 4 and 5 and phantoms 8 and 9 provided the same average particle diameter, so these phantoms were averaged to produce one simulation and experimental result per particle size. 

%For the simulation, the number density of the phantoms was averaged between the two phantoms of equal particle diameter which was equivalent to running the simulation for each number density and averaging the results. This exclusion and averaging produced a final four phantoms of varying particle sizes for analysis, summarized in Table \ref{table:2}. 
\begin{table}[ht]
\caption{Average particle size and volume density of the phantoms}
\begin{tabular}{|c|c|c|c|}
    \hline
    \rule[-1ex]{0pt}{3.5ex} Phantom & Particle Diameter ($\mu$m) & Surface Density (particles/mm$^2$) & Volume Density (mm$^{-3}$) \\
    \hline
   \rule[-1ex]{0pt}{3.5ex} A & 115 & 3 & 10 \\
    \hline
   \rule[-1ex]{0pt}{3.5ex} B & 130 & 2 & 7 \\
    \hline
   \rule[-1ex]{0pt}{3.5ex} C & 180 & 3.5 & 8.5 \\
    \hline
   \rule[-1ex]{0pt}{3.5ex} D & 280 & 6.5 & 10.5 \\
    \hline
\end{tabular}
\label{table:2}
\end{table}

\subsection{Monte Carlo simulation results}
\begin{figure}[ht]
\begin{center}
\begin{tabular}{c}
\includegraphics[height=5.5cm]{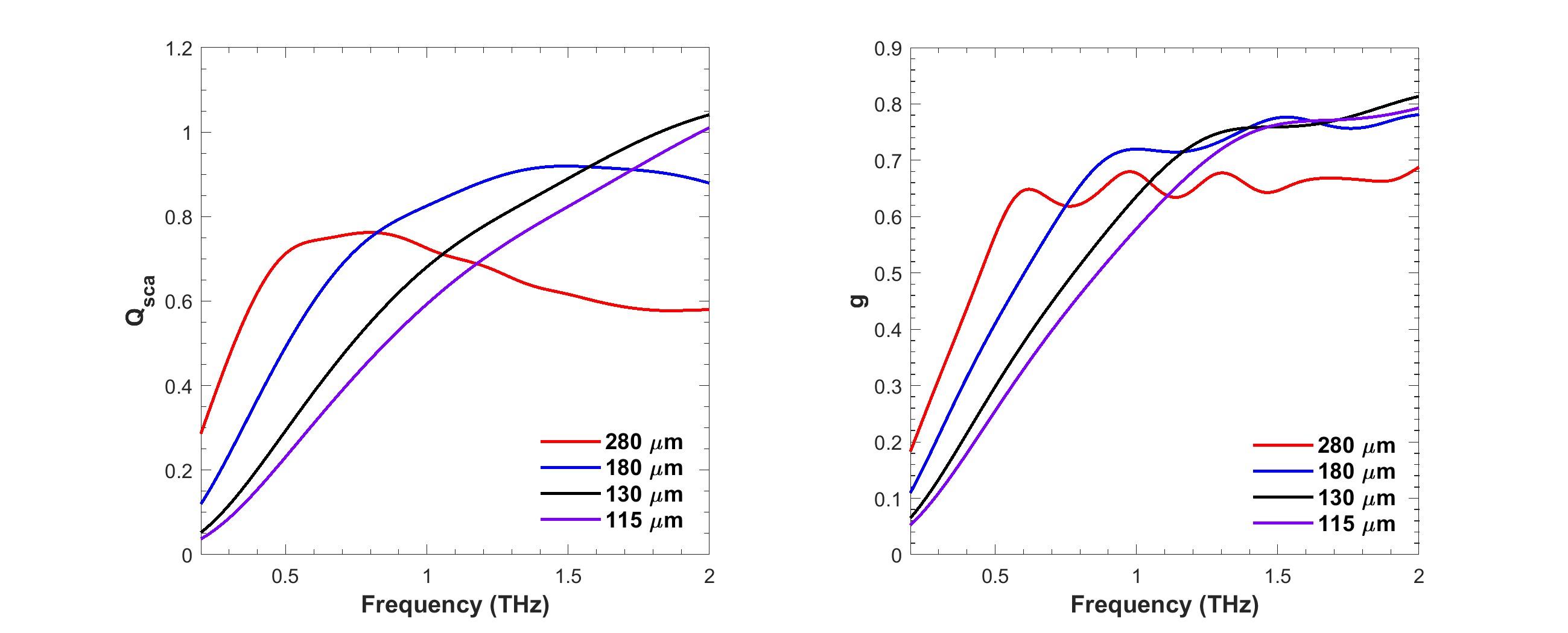}
\end{tabular}
\end{center}
\caption 
{ \label{fig:scatteringefficiency}
Calculated scattering efficiency and asymmetric factor for relevant particle sizes.
} 
\end{figure}

Using the refractive indices calculated in Section 2.3 and the particle sizes and volume densities determined in Section 3.1, the Polarized Light Monte Carlo model was used to simulated the THz polarimetric response of the final four phantoms. For each particle size, we calculated the asymmetric factor and the scattering efficiency, which is the ratio of the scattering cross-section to the geometric area of a particle. Figure \ref{fig:scatteringefficiency} shows the calculated scattering efficiency and asymmetric factor for each particle size. As the particle size increases, the scattering efficiency and asymmetric factor both increase in the lower frequencies up to about 1 THz. 

\begin{figure}[ht]
\begin{center}
\begin{tabular}{c}
\includegraphics{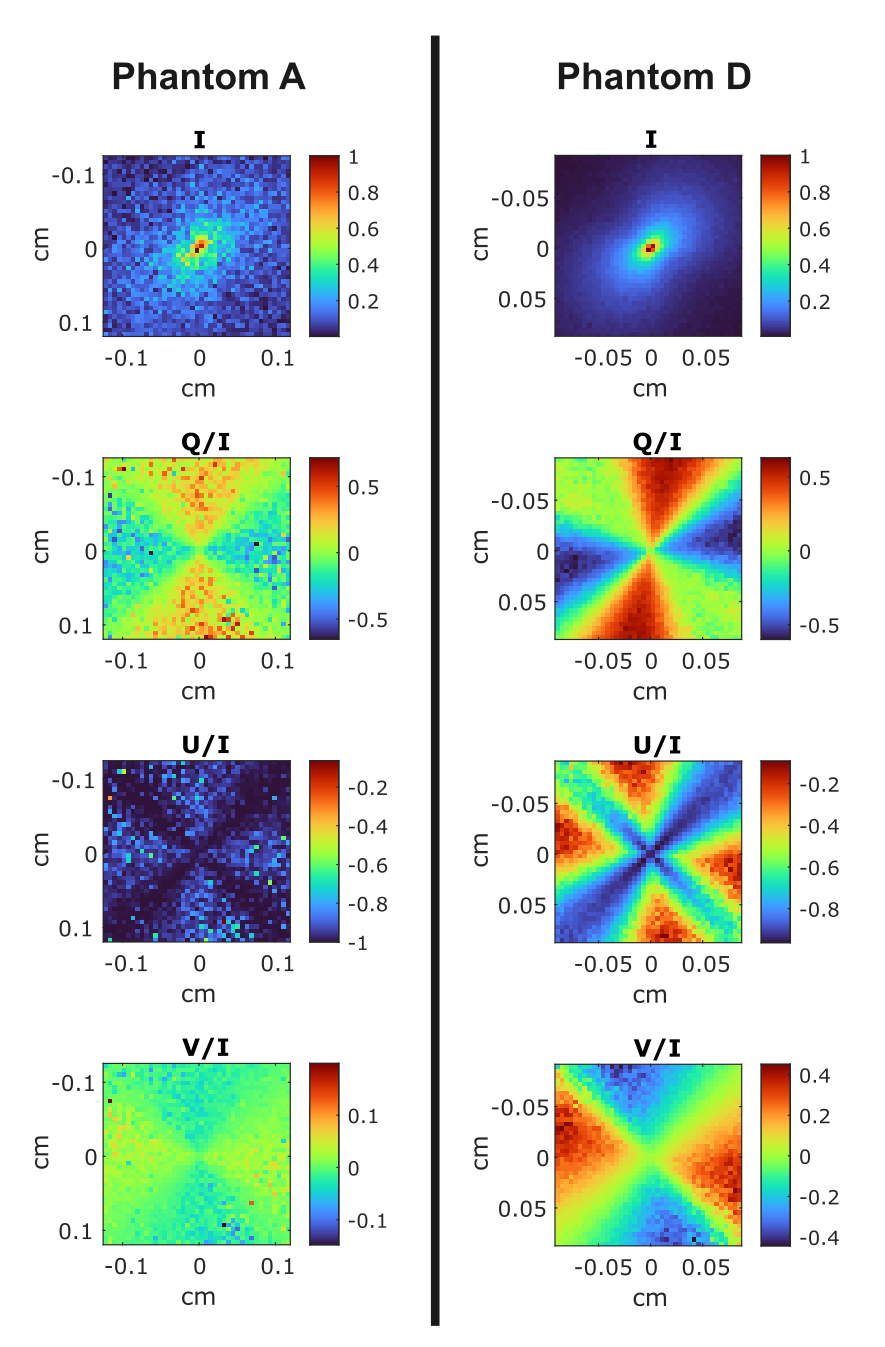}
\end{tabular}
\end{center}
\caption  
{ \label{fig:IQUV}
Spatial map of the simulated Stokes vector, IQUV, for Phantoms A and D.
} 
\end{figure}

\begin{figure}[ht]
\begin{center}
\begin{tabular}{c}
\includegraphics{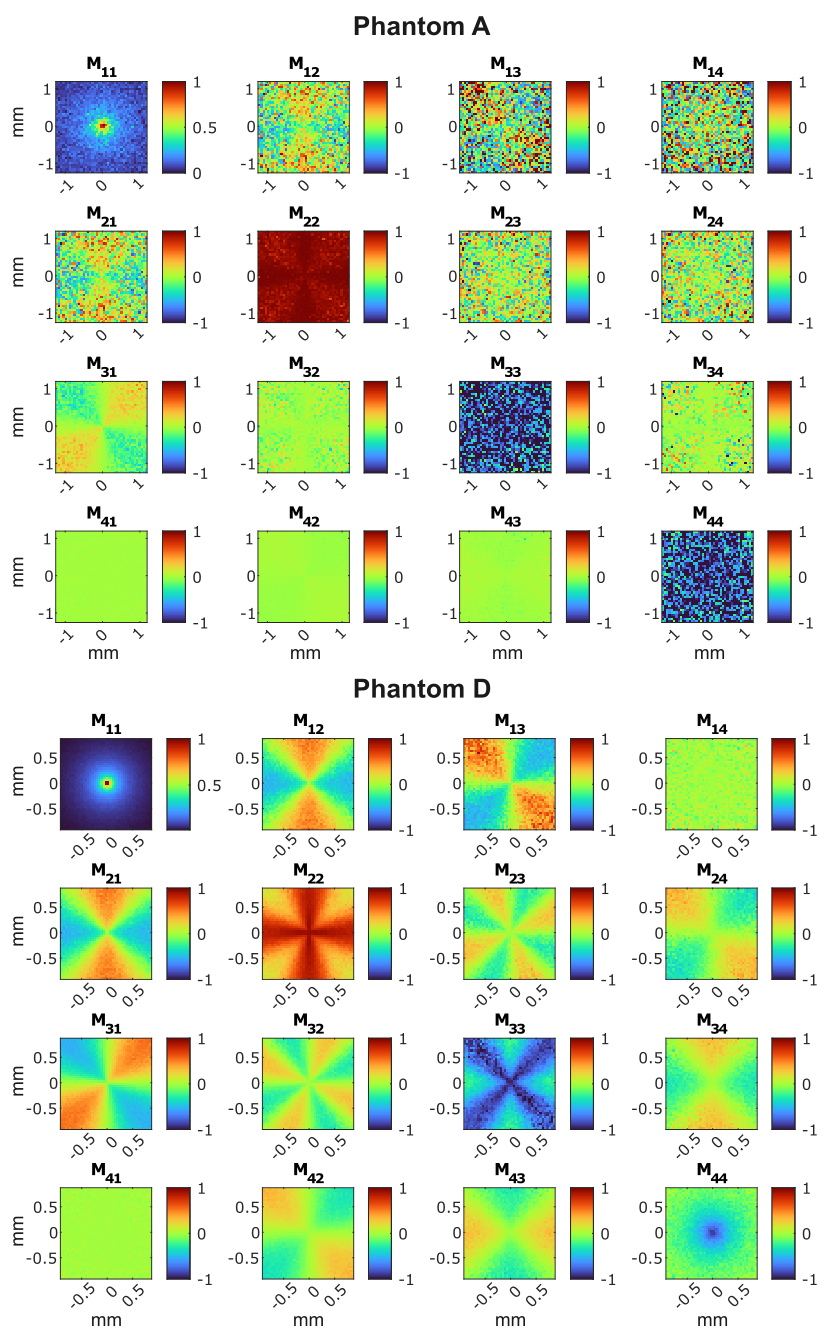}
\end{tabular}
\end{center}
\caption 
{ \label{fig:Mueller}
Simulated Mueller Matrix for Phantoms A and D
} 
\end{figure}

To calculator the Mueller Matrix of each sample, spatial Stokes vector maps were generated for each phantom for incident light polarizations of parallel, perpendicular, and 45\textdegree \space linearly polarized, and right circularly polarized. Figure \ref{fig:IQUV} presents a comparison between the Stokes parameters maps of Phantoms A and D, having the smallest and largest particle sizes respectively, when illuminated with a 45\textdegree \space linearly polarized incident THz beam. The spatial maps reveal the greater intensity of scattered light received with the simulation of Phantom D compared to Phantom A, matching the expectations of a higher reflectance with a larger particle size. 
%The spatial map of Q shows how with Phantom D, the distribution stays along the 45\textdegree \space angle, while with Phantom A, the distribution becomes more evenly distributed along all angles. Since there are less photons being reflected from the phantom with a smaller particle size, the spatial patterns are much less resolved with Phantom A compared to Phantom D. 
Using the spatial maps from all four incident light polarizations, the spatial Mueller matrix maps, shown in Fig. \ref{fig:Mueller}, were calculated. The simulated Mueller matrices for Phantoms D and A show similar angular distribution patterns, with the patterns being much less resolved with Phantom A, similar to the Stokes parameter maps. The clearest difference between the two maps is present in element $M_{22}$ where the values of the angular distribution pattern of Phantom D are closer to zero than the values of Phantom A. This difference is also reflected in element $M_{33}$ where the values of Phantom A are much closer to -1 than Phantom D. To better analyze the differences between the simulated Mueller matrices of the phantoms, we calculated the pixel sum of each of the Mueller matrix element maps to obtain a singular Mueller matrix for each phantom. The Mueller matrices of phantoms A and D, i.e., $M_A$ and $M_D$, are given below. 

\begin{equation}
    M_A = 3.05\times10^{-4} 
    \begin{bmatrix}
        1.000 & -0.003 & 0.008 & 0.008\\
        -0.003 & 0.922 & 0.004 & 0.004\\
        -0.002 & -0.001 & -0.926 & 0.000 \\
        0.000 & 0.000 & 0.000 & -0.850
    \end{bmatrix}
\end{equation}
\begin{equation}
    M_D = 4.23\times10^{-3}
    \begin{bmatrix}
        1.000 & 0.000 & -0.001 & -0.001\\
        0.000 & 0.586 & 0.000 & 0.000\\
        0.000 & 0.000 & -0.586 & -0.001 \\
        0.000 & 0.000 & 0.000 & -0.207
    \end{bmatrix}
\end{equation}
\par
These matrices can be decomposed using Lu and Chipman decomposition \cite{lu_interpretation_1996} as:
\begin{equation}
    M_{\Delta,A}\approx
    \begin{bmatrix}
        1 & 0 & 0 & 0 \\
        0 & 0.921 & 0 & 0 \\
        0 & 0 & 0.925 & 0 \\
        0 & 0 & 0 & 0.849 
    \end{bmatrix},
    M_{R,A}\approx
    \begin{bmatrix}
        1 & 0 & 0 & 0\\
        0 & 1 & 0 & 0\\
        0 & 0 & -1 & 0\\
        0 & 0 & 0 & -1
    \end{bmatrix},
    M_{D,A}\approx3.050\times10^{-4}\textbf{I}
\end{equation}

\begin{equation}
    M_{\Delta,D}\approx
    \begin{bmatrix}
        1 & 0 & 0 & 0\\
        0 & 0.586 & 0 & 0\\
        0 & 0 & 0.586 & 0\\
        0 & 0 & 0 & 0.207
    \end{bmatrix},
    M_{R,D}\approx
    \begin{bmatrix}
        1 & 0 & 0 & 0\\
        0 & 1 & 0 & 0\\
        0 & 0 & -1 & 0\\
        0 & 0 & 0 & -1
    \end{bmatrix},
    M_{D,D}\approx4.230\times10^{-3}\textbf{I}
\end{equation}

Each of the phantom's Mueller matrices can be decomposed as approximately a uniform attenuator and nonuniform depolarizer. Since this simulation records the reflected photons instead of transmitted photons through the medium, the diagonal value of the uniform attenuator matrix represents the diffuse reflectance of the sample. The diagonal depolarizer matrix can be represented with single value: the degree of polarization of linear light "p", where the depolarization matrix is diag[1, p, p, 2p-1]. Since the Mueller matrix and its decomposition can be represented in terms of only the diffuse reflectance and linear degree of polarization, only these two values are necessary to characterize the polarimetric response of these sample. These two parameters can be obtained by a measurement of either linear or circularly polarized light scattered to a diffuse angle; therefore, measurement of four different initial polarization states as would typically be done for a full Mueller matrix characterization with spherical symmetry is unnecessary in this case, according to this simulation. Using the spatial maps generated from linearly polarized light, we calculated the diffuse reflectance as the Stokes parameter I and the degree of polarization using Equation 5 to compare with experimental results.

\subsection{Comparison between model and experimental data}
For each phantom, time domain THz-TDS waveforms were converted into the complex fourier domain to produce $E_x$ and $E_y$ spectra. Stokes parameters were calculated using,
\begin{equation}
    I = \langle | E_x |^2 \rangle + \langle | E_y| ^2\rangle, 
    Q = \langle | E_x |^2 \rangle - \langle | E_y| ^2\rangle, 
    U = 2\Re\langle E_x E_y^* \rangle, 
    V = -2i\Im\langle E_x E_y^* \rangle.
\end{equation}
\par
\begin{figure}[ht]
\begin{center}
\begin{tabular}{c}
\includegraphics[height=9.5cm]{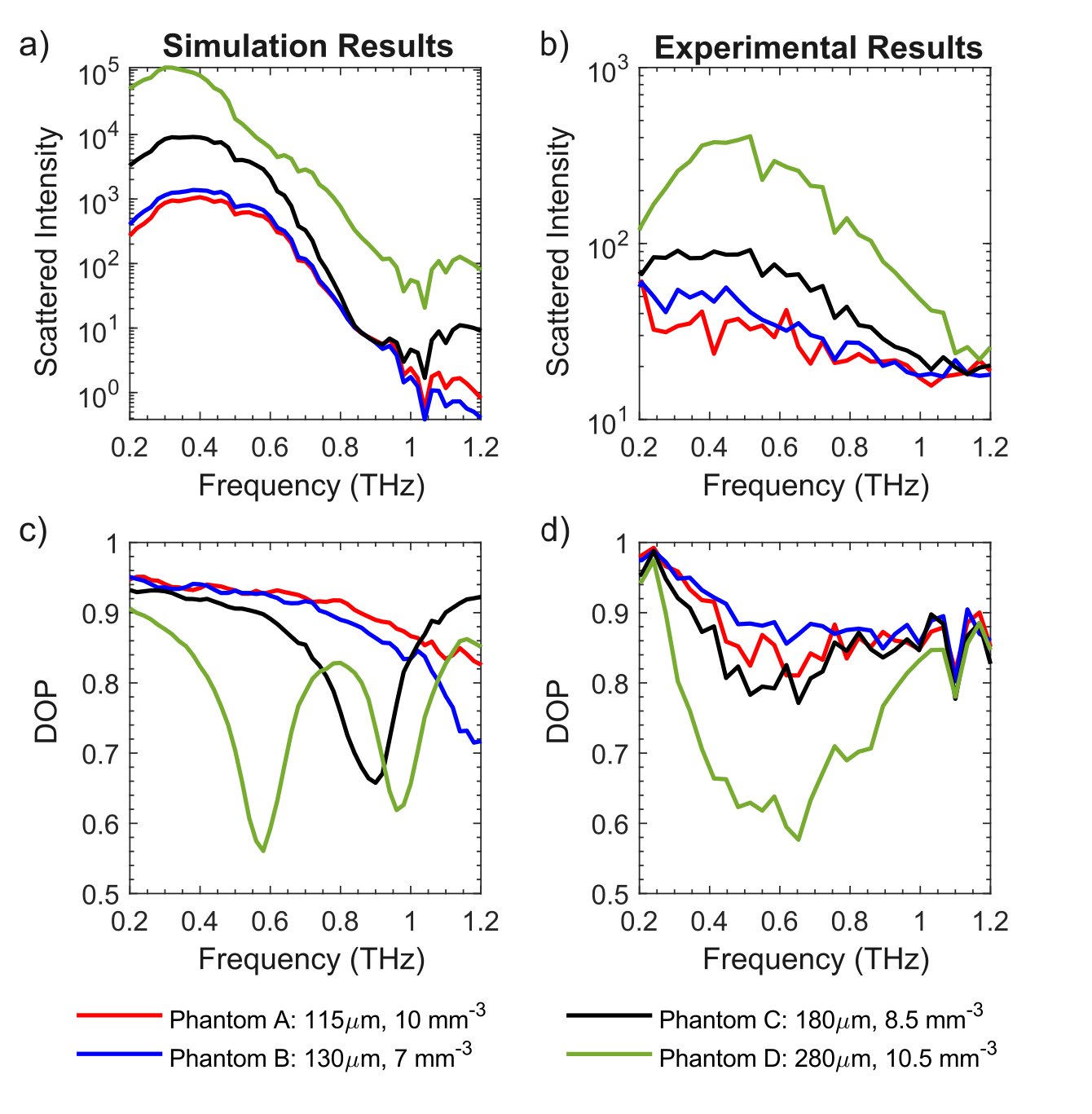}
\end{tabular}
\end{center}
\caption 
{ \label{fig:comparison}
Comparison between simulation and experimental data.  } 
\end{figure}

The intensity was taken as stokes parameter I averaged over 100 pixels, while the degree of polarization (DOP) was calculated using Equation 5 with the stoke parameters averaged across 100 pixels with $E_x$ and $E_y$ deconvolved with an air reference taken at 0 degrees transmission. The simulated result for the diffuse reflectance was converted into reflected intensity by multiplying the simulation by the spectrum of the air reference. Figure \ref{fig:comparison} presents the simulated results for linearly polarized light alongside the experimental results for the four phantoms analyzed. The diffuse scattered intensity between 0.2 and 1.2 THz has a similar trend between the simulation and experimental results, show a higher intensity at lower frequencies and an exponential roll-off with increasing frequency. As the diameter of the particles increases, the scattered intensity becomes greater. The absolute values of the diffuse scattered intensity between the simulation and experiment cannot be directly compared because the simulation results represents photons from all backscattered angles, whereas the experimental data was collected only in a 20\textdegree detection cone. The diffused scattered intensity is also very sensitive to slight changes in the alignment of the experimental system.

\par

The DOP parameter, which is less susceptible to the miss-alignments in the system, shows a similar trend to the diffuse scattered intensity results. Due to the SNR and bandwidth limitations of our system, DOP measurements past 0.8 THz are not as reliable %and are thus shaded gray in Fig. \ref{fig:comparison}.
As shown in Fig. \ref{fig:comparison}, phantom D shows a similar trend in both simulation and experimental data with a drop in DOP to around 0.6 at roughly 0.6 THz. Phantom C has less of a drop in DOP from 0.2-0.8 THz than Phantom D, but more of a drop than Phantoms A or B. Phantoms A and B have an opposite trend than expected, but this reveals the limitations of our system. Since the frequency location of the drop in DOP depends on particle size, to measure a drop in DOP from smaller particle sizes, a larger bandwidth than our experimental setup would be required. Despite the lack of agreement with Phantoms A and B, there is still a significant difference in measured DOP results between Phantom D and Phantoms A and B. Additionally, DOP is much less sensitive to the alignment of the setup, so it could be a more reliable diagnostic marker than reflected intensity.

\subsection{Experimental Data of A Porcine Skin Burn}
\begin{figure}[ht]
\begin{center}
\begin{tabular}{c}
\includegraphics[height=9.5cm]{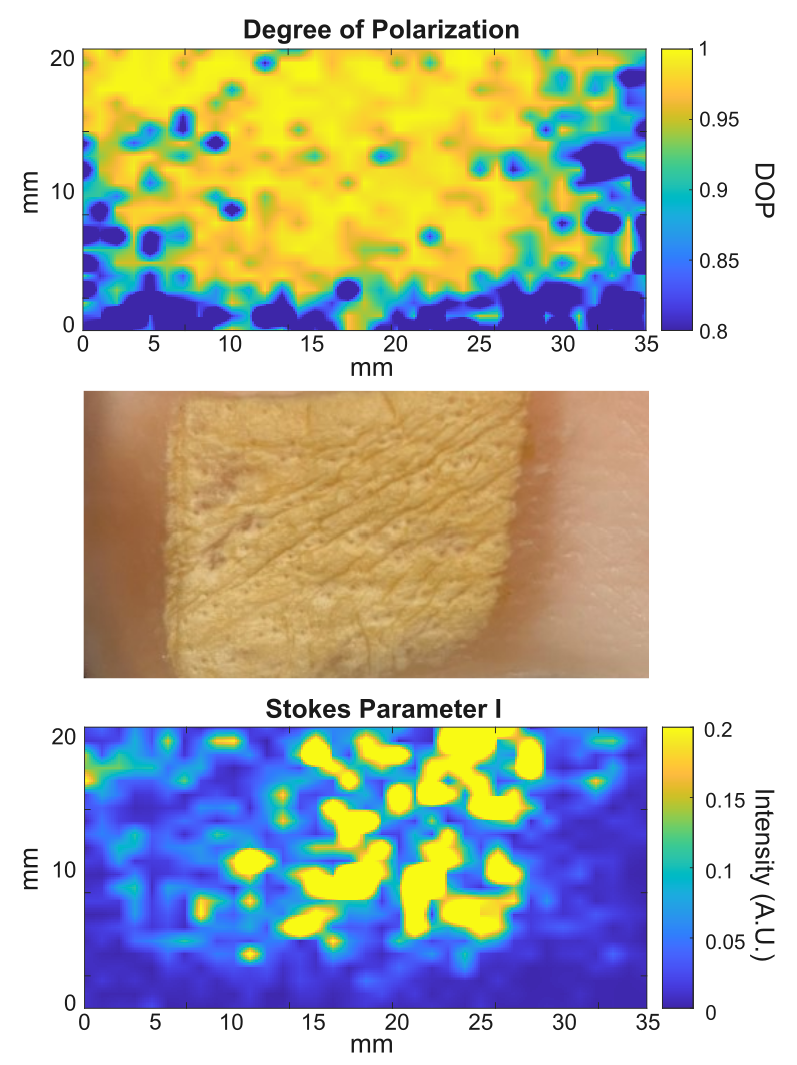}
\end{tabular}
\end{center}
\caption 
{ \label{fig:burnmaps}
DOP and diffused scattered intensity maps of a porcine skin burn at 0.6 THz. } 
\end{figure}

To demonstrate the potential of DOP and diffuse backscattered intensity in realistic tissue conditions, we present results of THz measurements of an ex vivo porcine skin burn. A 1-inch full thickness burn was induced by applying a brass bar heated to 180\textdegree C to the skin for 5 minutes. We mounted the porcine skin in our setup at 140\textdegree \space and took measurements of a 2 x 3.5 cm area with a 1mm pixel size. Each pixel had 10 measurements taken with 10 time-averages. Using the 10 measurements per pixel, we calculated the intensity as the Stokes parameter I and the DOP using Eqns. 5 and 6. Figure \ref{fig:burnmaps} presents spatial maps of the DOP and intensity at 0.6 THz compared with a photograph of the burn area imaged. Both maps show contrast between the burn and healthy tissue. A higher DOP values for the burn could be attributed to the destruction of large skin structures such as hair follicles and sweat glands, overall decreasing the average size of scatterers within the skin. The higher intensity values for the burn could be attributed to decreased absorption from the skin due to water loss from the heating of the skin in the production of the burn. 

\begin{figure}[ht]
\begin{center}
\begin{tabular}{c}
\includegraphics[height=9.5cm]{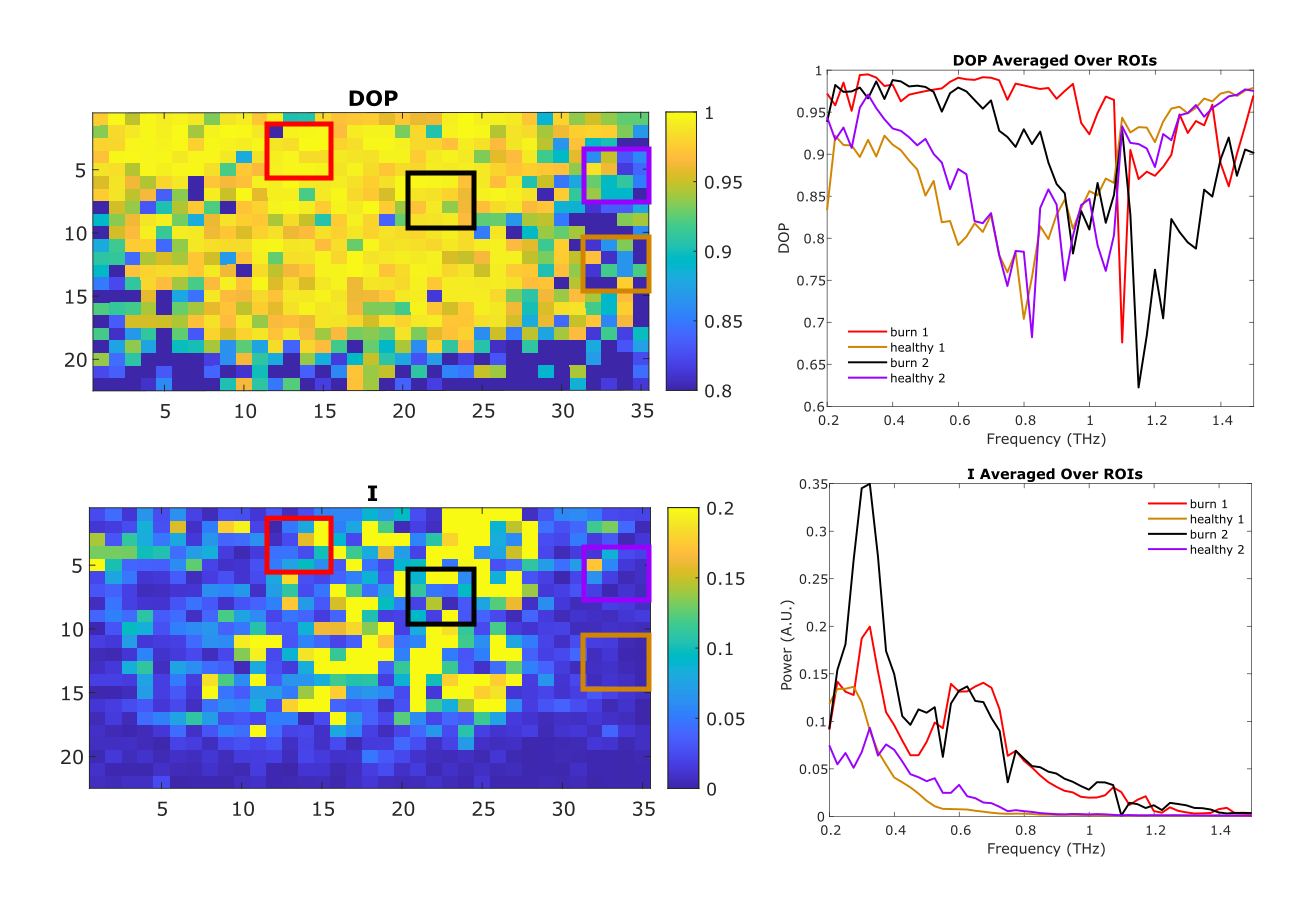}
\end{tabular}
\end{center}
\caption 
{ \label{fig:burnrois}
Frequency-dependent DOP and Intensity of two burn ROIs and two healthy porcine skin ROIs. Each ROI is composed of 16 pixels. } 
\end{figure}

To compare the frequency-dependent burn measurements with the healthy tissue measurements, four ROIs were averaged and plotted, two burn ROIs and two healthy tissue ROIs, each with 16 pixels. Figure \ref{fig:burnrois} shows these frequency-dependent curves and their locations on the THz maps. These curves reveal that the healthy tissue DOP drops much faster than the burn DOP from 0.2-0.8 THz. According to our simulation data, this would indicate a decrease in the size of scatterers within the burn tissue compared to healthy tissue. We also observe a large contrast in the diffuse intensity between the burn and healthy tissue. According to the Mie scattering simulation, an increase in DOP should coincide with a decrease in diffuse intensity, but the opposite pattern is seen with this burn. This could be attributed to a difference in absorption coefficient since there is a decrease in water content of the tissue after a burn injury is inflicted. This would increase the diffuse intensity since the overall albedo would increase. 

\section{Discussion}
\begin{figure}[ht]
\begin{center}
\begin{tabular}{c}
\includegraphics[height=5.5cm]{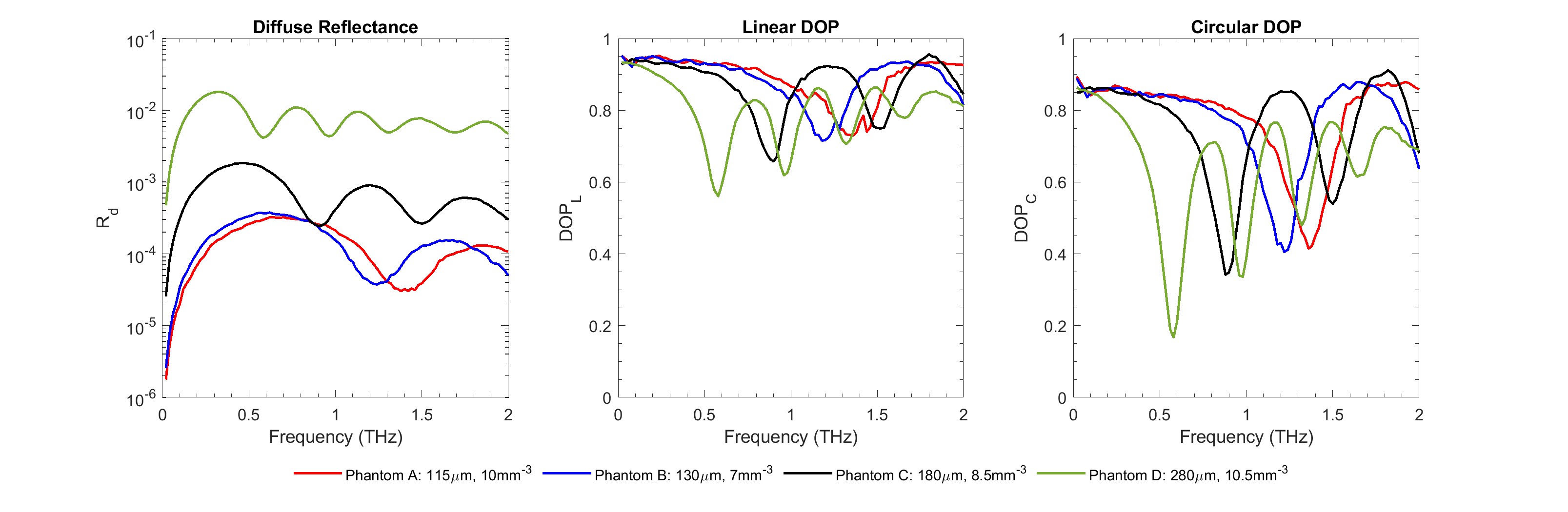}
\end{tabular}
\end{center}
\caption 
{ \label{fig:rdanddop}
Simulated Diffuse Reflectance and DOP up to 2 THz.
} 
\end{figure}

Extending the simulation results to a larger bandwidth of 2 THz (Fig. \ref{fig:rdanddop}) shows more clearly the frequency-dependent features of the DOP. These results reveal that as the particle size increases, the diffuse reflectance increases. Additionally, the spectral location of the first minima in the DOP (for both linearly and circularly polarized light) moves to lower frequencies as the particle size increases. Xu and Arbab showed in their analysis of this simulation method that the depth of the minima in  DOP curves is a function of the concentration of the particles. In other word, a larger drop in DOP corresponds to a higher concentration \cite{Xu:24}; therefore, spectral dependence of the DOP can allow for determination of both the concentration and size of particles, given a frequency-dependent DOP measurement with sufficient signal-to-noise ratio and bandwidth. The circular DOP ($DOP_c$) is related to the linear DOP ($DOP_l$) by the relationship $DOP_c = 2DOP_l-1$. Finally, the magnitude of the diffuse reflectance changes with both particle size and concentration. Our results suggest that the particle size-dependent pattern in the diffuse reflectance may be difficult to resolve with current THz systems. 

\par
Given the results of our phantom measurements and the diffused polarimetric imaging of a porcine skin burn, contrast can be seen with the DOP and diffuse scattered intensity between different sizes of scattering particles in tissue environments; however, these signal contrasts are best observed when the difference between the sizes of particles are large. Observations of smaller differences in concentration and size of scattering particles are limited by our current bandwidth and SNR. A THz system with a larger bandwidth, such as the one developed by Xu et al. with a bandwidth of 8 THz using electro-optic crystals instead of PCA emitters and detectors \cite{xu_broadband_2022}, could aid in resolving these features for smaller particle size differences. 

Lastly, in this work we assumed that the surface of the phantom and the ex vivo skin is optically flat and smooth.  We have previously studied the depolarization of THz waves in the backscattered speckle fields due to surface roughness of the sample \cite{Xu:23}. In general, the DOP parameter can be spectrally affected by both surface depolarization as well as scattering by the internal structures of the skin. Several signal processing and experimental measurement techniques can be used to characterize the relative contribution of each source of depolarization. Examples of these methods include wave signal decomposition \cite{8509945,khani:22:se} and statistical speckle analysis \cite{Xu:23}.

\section{Conclusion}
We have presented simulation and experimental results for Mie scattering of THz light due to structures in biological tissue, such as hair follicles, sweat glands and tumor budding. Simulations revealed the potential for diffuse backscattered intensity and Degree of Polarization (DOP) parameter to distinguish between different scattering scenarios. Simulated Mueller matrices showed that for spherical particles in an absorbing medium, only one polarization measurement from either linearly or circularly polarized light is needed to construct a full Mueller matrix of the sample. Additionally, we decomposed the Muleler matrices of the tissue according to the Lu-Chipman formalism and summarized the key sources of the scattering signal contrast in highly absorptive media. Experimental measurements of phantoms of moderately sized tumor budding and poorly differentiated clusters confirmed the frequency-dependent patterns from our simulation and showed the potential of DOP and diffuse scattered intensity for diagnosis. Finally, experimental measurements of a porcine skin burn showed contrast between burned regions and healthy regions of tissue, demonstrating the ability of this technique to distinguish between disease states in ex vivo tissue. Future work in this direction would include extending our phantom measurements to ex vivo cancer tissue, increasing the bandwidth of the measurements to better resolve the frequency-dependent features.

%\subsection* {Acknowledgments}
%This unnumbered section is used to identify those who have aided the authors in understanding or accomplishing the work presented and to acknowledge sources of funding. 

%%%%% References %%%%%

\bibliography{article}   % bibliography data in report.bib
\bibliographystyle{spiejour}   % makes bibtex use spiejour.bst

%%%%% Biographies of authors %%%%%

%\vspace{2ex}\noindent\textbf{First Author} is an assistant professor at the University of Optical Engineering. He received his BS and MS degrees in physics from the University of Optics in 1985 and 1987, respectively, and his PhD degree in optics from the Institute of Technology in 1991.  He is the author of more than 50 journal papers and has written three book chapters. His current research interests include optical interconnects, holography, and optoelectronic systems. He is a member of SPIE.

\vspace{1ex}
%\noindent Biographies and photographs of the other authors are not available.

\listoffigures
\listoftables

\end{spacing}
\end{document}